\documentstyle[12pt]{article}
\newcommand{\bcn}{\begin{center}}

\newcommand{\beq}{\begin{equation}}

\newcommand{\beqn}{\begin{eqnarray}}

\newcommand{\ecn}{\end{center}}

\newcommand{\eeq}{\end{equation}}

\newcommand{\eeqn}{\end{eqnarray}}

\def\appgt{\mathrel{\mathpalette\@versim>}}
\def\@versim#1#2{\lower2pt\vbox{\baselineskip0pt \lineskip-.5pt
   \ialign{$\m@th#1\hfil##\hfil$\crcr#2\crcr\sim\crcr}}}
\begin{document}
\baselineskip 18pt
\voffset=-0.6in
\rightline{KANSAS 94-5-22}
\rightline{OKHEP-94-02}
\bigskip
\bcn{\large\bf Evidence for Gluon Energy Loss as the Mechanism for Heavy
Quarkonium
Suppression in  $pA$ Collisions\\}
\bigskip

{\large\bf Pankaj Jain}\\
\smallskip

{\it Department of Physics and Astronomy\\
University of Oklahoma, Norman, OK 73019\\}
\medskip
and\\
\medskip
{\large\bf John P. Ralston}\\
\smallskip
{\it Department of Physics and Astronomy\\
and\\
Kansas Institute for Theoretical and Computational Science\\
University of Kansas,
Lawrence, KS-66045}\\

\bigskip
\ecn

\noindent{\bf Abstract}

\medskip
\noindent
We study the energy and nuclear A dependence of the hadronic production
of heavy quarkonia.  We review theoretical ideas which have been put
forward, seeking a consistent global picture reconciling the large effects
in quarkonia with the small nuclear effects observed in continuum Drell
Yan production.  The data indicates that  shadowing or
leading twist modifications of parton distributions can be ruled out as
explanations, leaving higher twist energy loss.
{}From general principles the maximum allowed energy loss of partons traversing
the nuclear
medium can be related to the parton transverse momenta.  We then show
that the experimental data on nuclear suppression of charm- and bottom-
onium for large $x_F$ is consistent with this effect: using the observed
transverse
momenta to bound the $x_F$
dependence in an almost model independent manner generates a relation that
practically reproduces the data. Several prediction are
discussed; the dependence on $x_F$ as $x_F\to 1$, and large and small $k_T^2$
cuts, can be used to discriminate between quark and gluon induced effects.

\vfill
\eject

\noindent
{\bf 1. INTRODUCTION}

\medskip
Propagation of quarks, gluons and hadrons through nuclear matter is
currently a subject of intense interest. It is expected that the study will
teach us much about the interplay between perturbative and non-
perturbative QCD.  An important experimental discovery by the Fermilab
E772 experiment [1] is the suppression of charmonium and bottomonium
production in $pA$ collisions in comparison to the production rate in $pp$
collisions. A similar effect had also been previously reported by Badier et al
[2] and Katsanevas et al. [3].  The data strongly contradicts a widespread
theoretical
expectation that at high energies the nuclear medium should have
negligible effect on heavy quarkonium production.  The observed
suppression has direct implications for the use of onium production as a
signal for quark-gluon plasma formation in heavy ion collisions.  It has
generated much controversy [4-7], and raised the possibility that  the
observations represent a serious challenge to theory.

\medskip
A common theoretical prejudice that suggests negligible nuclear
suppression is the following.  One can argue that the characteristic
separation of the charmed quark anti-quark pair produced by a partonic
interaction is of the order  of $1/m_c$, which is equal to (1.5 GeV)$^{-1}$.
Invoking a geometrical cross section in the spirit of color transparency,
the attenuation cross section of the charm pair might be as small as the
order of (1.5 GeV)$^{-2}$ , which is about 0.2 mb.  With this cross section,
a typical survival factor of 0.97 is obtained for a large nucleus of
diameter 10 fm.  This is a very small effect in comparison to 30 to 60 \%
suppression seen in the data.

\medskip
Of course, such a method of estimating the nuclear effects applies (at
best) to the propagation of a well-localized, relativistic, color-singlet
charm pair.  But invoking color transparency for the actual production is
probably not correct, since the kinematics of the events are highly
inelastic, and lack the usual conditions of exclusivity that color
transparency arguments should assume.  Color transparency is not expected to
occur if the coherence of a system is broken, for example in the case
when uncontrolled inelastic color flows are summed over in a semi-inclusive
production.  Moreover, in the initial state the gluons can lose energy by
interacting with the nuclear medium, and there also is the likelihood that
the charmed pair is produced temporarily in the color octet (rather than
singlet) state.  This does not mean that the suppression is of the expected
size: indeed, the huge magnitude of the effect has been quite mysterious.

\medskip
Besides the outstanding puzzle of quarkonium suppression, the general problem
of parton
propagation in the nucleus is becoming rather important. There is great
interest in understanding the role of quantum mechanical coherence of QCD
interactions in the nuclear medium. Depending on the experimental
circumstances,
the same arguments leading to coherent suppression of energy loss in high
energy QED interactions - the so called LPM [8-10] effect - can be applied,
leading
to many interesting predictions. In the LPM effect, a concept of ``formation
time'' $\tau_{\rm form}$ of quanta during a high energy interaction is
considered. The order of magnitude of the formation time for a quantum of mass
$m$, carrying
energy $E$, and transverse momentum $k_T$ with respect to its progenitor is
\beq
\tau_{\rm form} \approx E/(k_T^2+m^2)
\eeq
The LPM effect in QED suppresses bremsstrahlung due to destructive interference
between emissions occuring over an interaction time $\tau_{\rm int}\le
\tau_{\rm form}$. A recent paper by Gyulassy and Wang [11] studies the effects
of
multiple
scattering in perturbative QCD. Although the details of coherence are
considerably more complicated due to the non-Abelian color algebra, the basic
features of the LPM effect are not changed by this study.

\medskip
The concept of the formation time leads to a complementary concept of formation
rate $\Gamma_{\rm form}$. The formation rate is simply the inverse of the
formation
time,
\beq
\Gamma_{\rm form} \approx (k_T^2+m^2)/E
\eeq
Examined in this way, there is an  interesting possibility that the coherent
parton formation rate could be ``tuned'' by selecting signals with various
masses, energies and transverse momenta. Is it possible, for example, to
increase
particle formation by selecting events with larger $k_T^2$? As we will show
below, the trends in the data indicate possible observation of a dramatic
``anti-LPM'' enhancement of parton emission due to increased formation rate
associated with large $k_T^2$. The possibility of tuning the formation rate
leads naturally to a number of interesting signals which can be experimentally
tested.

\vglue 0.3in
\noindent
{\bf 2. REVIEW}

\medskip
In view of these points (and the historical difficulty of understanding
charm production in general), we will approach the problem of onium suppression
in a
methodical manner, and try to limit our assumptions carefully:

\medskip
\noindent
i) Electroproduction experiments have convincingly shown that the
process of parton fragmentation is negligibly affected by nuclei.  The
E772 experiment itself also found that continuum Drell-Yan muon pair
production shows no significant nuclear effect.  The data for continuum
dimuon $A^{-1} d\sigma/dQ^2$, for example,  shows little nuclear effect in the
$Q^2$ regions both above and below the suppressed onium regions.  These
observations are consistent with conventional factorization.  They seem to
significantly limit the amount of energy loss one can attribute
phenomenologically to parton propagation inside the nuclear medium.

\medskip
However, we note that both the bulk of fragmentation and the Drell
Yan continuum are mainly probes of the quark-antiquark channel.  They say
little about production via gluons.  The nature of gluon and heavy quark
propagation in
nuclei is not clear.

\medskip
\noindent
ii)  A very important clue to the charmonium data is given by the fact that
the data's $x_2$ dependence does not scale.  Even if one creates an ad-hoc
gluon
or quark-antiquark distribution for one experiment on charm, it does not
reproduce the data for bottom. Furthermore the data does not scale with $x_2$
as we change the incident beam momentum from 200 to 800 GeV [12].  This
eliminates the possibility of gluon
shadowing or more general EMC-type effects on the parton distributions
as a dominant mechanism.  Quite recently, Banesh, Qiu and Vary [7] (BQV) claim
that the onium production at large $x_F$ is dominated by quark annihilation,
and
that shadowing in the small-$x_2$ quark distributions  accounts for the
dominant part of the suppression. Of course, shadowing makes a contribution,
 but the
magnitude of small $x$ shadowing observed in deeply inelastic scattering is
not nearly large enough to account for the onium suppression. Postponing a
detailed discussion to Section 5, we do not believe that the BQV analysis
actually support
the claim. The same
objection based on factorization must be raised here: the onium production
data shows that factorization does not hold,  ruling out this mechanism
as a model for the full effect.

\medskip
\noindent
iii) Since factorization is a leading twist prediction for this kind of semi-
inclusive production, and the same sort of experiment has confirmed its
use and understanding elsewhere, the suppression effect must be a higher
twist one.  The bizarre thing is that it so drastically affects the onium
production.

\medskip
\noindent
iv) The quarkonium suppression seen in the E772 data can be questioned as
to whether it might be an instrumental effect.  In fact, certain regions of
the data's transverse momentum spectra have not been reported due to
questions about the acceptance [13].  We re-examined previous data of Badier
et al [2] on $J/\psi$ production on nuclear targets in studying this.  We find
that
rather than contradicting the E772 experiment, the Badier et al data
seems to confirm the trend.  A signal for suppression of onium is real and
more than ten years old.

\medskip
We now summarize some theoretical ideas, which we consider as "spare parts"
that might be
assembled in a new order to understand the puzzle:

\medskip
\noindent
a) It can be argued that heavy quarkonium, as a non-relativistic
bound state, might have a large nuclear matter interaction
cross section that is set by the full onium size and binding energy rather
than the mass.  The $J/\psi$ bound state diameter, for example, is 3 to 5 times
the charmed quark compton wavelength, leading to an estimated cross
section as large as 2 to 5 mb.  A major problem with this proposal is that the
time scale for formation of a bound state onium is very long compared to
the time scale for crossing the nucleus.  This is aggravated by the fact
that the Lorentz boost stretches any effective time scale enormously, so
that a $J/\psi$ does not become a $J/\psi$ until it is more than 100 Fm away!
 We do not pursue this idea further.

\medskip
\noindent
b) There remains a realistic possibility that interaction of incoming
and outgoing colored partons could cause them to lose energy.  Gavin and
Milana (GM) [5] observed that even a small shift in the $x_F$ value of the
incoming partons, assumed by them to be gluons, could lead to a
numerically large change in $d\sigma/dx_F$ because of the rapid variation
with $x$ of the gluon distribution functions.

\medskip
For example, supposing the gluon distribution to go like $(1-x)^5$, then a
gluon taken from the gluon distribution at $x + \delta x$ is less likely to
contribute by the ratio $(1-x-\delta x)^5/ (1-x)^5 \approx 1 - 5 \delta x
/(1-x)$.  Even if $\delta x$ is
small, this creates a strong kinematic suppression as $x$ approaches unity.
For reference we will call this the GM mechanism.  This effect was
assumed to occur for both initial and final state propagation, with onium
production dominated by color octet components.

\medskip
To get a large enough shift in the $x$ value, GM also proposed a
rule for energy loss which is of the ``higher twist'' type; their proposal for
the energy loss is  $\Delta E/E=  c x_1 A^{1/3}/Q^2$, where c is a
color dependent constant that could be adjusted to fit to the data .  At
first the higher twist character of the GM rule seemed to put it into a
phenomenological limbo of the uncalculable, a thing which could be neither
verified nor disproven with current theoretical knowledge.

\medskip
\noindent
c) This approach was countered by Brodsky and Hoyer (BH) [6] who
argued that the dependence on energy of the GM formula
violates general principles.  BH went on to claim that there exists an
upper bound on the energy loss for a parton propagating through nuclei.
This upper bound, in the spirit of LPM, is obtained from rather general
considerations; the BH
rule is $\Delta x < k_{_T}^2L_A/2E$, where $k_{_T}$  is the transverse
momentum change in the collisions giving the energy loss and $L_A$ is the
target length.  This rule also
is higher twist, but the uncalculable higher twist mechanism of Gavin and
Milana has a limit that contradicts this relation, allowing $\Delta E$ to go
like $E$ at fixed $Q^2$.  The contradiction between the two formulas is
numerically important.  Using their own bound for the energy loss and a
value for $k_{_T} $= 300 MeV, Brodsky and Hoyer dismissed the Gavin and
Milana proposal as insufficient to explain the data.

\newpage
\noindent
{\bf 3. ASSEMBLING THEORETICAL SPARE PARTS}

\medskip

Considering these ideas, let us first
observe that the GM mechanism is (a) quite reasonable and (b) logically
independent of the model used for the energy loss.  A first task, then, is to
determine
whether some energy loss, however it occurs, can explain the data.  This
has already been answered by GM who were able to fit the data.  There is
no reason, then, to rule out gluon energy loss as a mechanism, but we agree
that
one should have a consistent framework to represent it.

\medskip
A second observation is that the BH expression for energy loss can
be viewed a testable hypothesis, that is, as a model.  The rule is claimed
to be an upper bound, so any observed energy losses must lie inside an
envelope of values it specifies.  This prompts us to examine the
experimental data, using the GM hypothesis, and test for the
general mechanism of energy loss by seeing whether or not the data obeys
the claimed bound.  Our goal here is to check the energy loss proposal without
getting snarled into model dependence of the energy loss formula.

\medskip
Third, given the fact that Nature tends to dissipate energy rather
maximally, we can try saturating the bound and examining whether we
obtain a prediction close to the behavior shown in the data.  Actually, by
examining the bound more closely, we find it needs modification from mass
effects and also by an unknown
dimensionless factor.  We consider our additions to be modest corrections.  The
trends in the data are not
strongly dependent on the unknown prefactor.

\medskip
The arguments leading to the BH bound, its modifications,
and its application along with the GM mechanism are reviewed in the next
section. We find that we can test for energy loss by examining the
detailed $x_F$ distribution using the $k_T$ distribution as an input.  This
is a much more powerful test than simply looking at overall production rates.
The
procedure works quite well in the upsilon meson case where the $k_T$
distribution has been measured.  We find that the suppression seen in the
$x_F$ distribution is well within the limits imposed by the bound on
energy loss. Even more importantly, the dependence of the data on $x_F$ and $A$
actually tend to parallel the bound.  We consider the agreement of a general
bound and a hitherto
unnoticed pattern in the data itself to be practically model independent
evidence that the basic culprit in nuclear quarkonium suppression is gluon
energy loss.

\medskip
While the bottom quark case checks well,
a definitive application of this result to E772 charmonium
production is not yet possible because of problems due to experimental
acceptance [13].  The transverse momentum distribution for the charmed
case is
known only for $k_T$ less than about 2 GeV.  We can nevertheless apply
our formalism in reverse and use the experimentally observed $x_F$
distribution to put a lower limit on the $k_T$ distribution of charmonium.
This prediction can be tested in future experiments.  Our results show
that the lower limit on the $k_T$ distribution is roughly the same as
the corresponding observed distribution for the case of bottomonium.

\vglue 0.3in
\noindent
{\bf 3.1 THE ROLE OF GLUONS}

\medskip
Although we will present evidence that the GM mechanism is at work, this
connection seems,
at first sight, not specific to the production mechanism.
There is a problem due to the generality of the BH bound.  The BH bound
assumes too little; one might say it tells us that we are seeing the
uncertainty principle at work, a fact of limited usefulness.  One knows
energy was lost but does not identify the underlying subprocess with this
mechanism alone.

\medskip
 The surprise of the onium suppression is contained in the large
magnitude of the effect.  One of our main points is that the quarkonium
suppression (which has gotten so much attention) is directly related to the
transverse momentum in the data (which has gotten very
little attention).  The average transverse momentum squared $<k_{_T}^2>_A$
is rather large;\footnote{A similar observation on the largeness of nuclear
interaction induced tranverse momentum in dijet production has been made by T.
Fields and M. Corcoran (to appear in {\it Proceedings of EPS Conference},
Marseille 1993). We thank Tom Fields for informing us of this.} for Tungsten
the $<k_{_T}^2>_A$ for bottomonium is greater than 2 GeV$^2$
rather
than 0.1 GeV$^2$. As a function of A, the integrated $\Upsilon_{1s}$
data is well described by [1,14]
\beq
<k_{_T}^2>_A  = \left[0.16 (A^{1/3}-2^{1/3}) + 2.59\right] \ {\rm GeV}^2\ \ .
\eeq
The same experiment found the $A$ dependent part of the Drell Yan continuum
$<k_{_T}^2>_A$ to be
about 10 times smaller.  As we will show, much can be predicted simply
knowing the $<k_{_T}^2>_A$.

\medskip
What, then, is causing the rapid variation with $A$ of the quarkonium
$<k_{_T}^2>_A$? Variation of mean transverse momentum with $A$ has also been
seen in other
experiments studying nuclear dependence of dijet and dihadron production
[15,16].
Physically this effect can arise from multiple scattering in the nuclear medium
[17]. The data's $A^{1/3}$
dependence indicates a target length effect, consistent with the energy
loss mechanism.  The same data indicates that the charged partons
contributing to Drell Yan are acting different from the partons making the
heavy quarkonia.  Drell Yan precludes any strong effects for the quarks,
then! This suggests that gluons, which give dominant contribution to
heavy quarkonia production, are
 scattering more in the transverse direction and are losing more energy than
quarks. An alternate possibility, which is not strictly ruled out by the
current data,
is that the final state interactions of the
heavy quarks are responsible for the large nuclear effects. However, experience
with QED bremsstrahlung and perturbative QCD, where a quark mass acts to
substantially cut--off massless vector emissions, makes it hard to believe that
heavy quarks could lose energy so much faster than light quarks!
In Section 5 we show that analysis of
data at large $x_F$ will also be able to discriminate clearly between these two
possibilities.

We believe that further data and not theoretical arguments will play the most
important role in finally pinning down all the unknowns.

\medskip
For better or worse, the dynamics of gluon channels producing onium
have never been clear, and a number of issues, including ``intrinsic charm'',
have been created to address the problem.  One cannot assign a
perturbative overall normalization to the gluon channels safely.  It is not
known, e.g. what fraction of the time a color octet $q\bar q$ state is
produced rather than a color singlet.  For this reason, we have arranged our
calculations so as not to base them on normalization factors. We
will postpone to
Section 5  a discussion of the interplay of subdominant quark
channels with the gluon channels.

\vglue 0.8cm
 \noindent
{\bf 3.2 ENERGY LOSS RATES}

\medskip
We first review and expand on the argument of Ref. [6] to obtain an
expression for an upper bound on energy loss. Assume a parton propagating
through the nuclear medium in the +z direction with energy $E=x_1 E_p$,
where $E_p$ denotes the energy of the incoming proton. This parton loses
an energy $x_g E$ by emitting a gluon in the presence of a source as shown
in Fig. 1.  This, and any secondary scattering, must occur in the volume of
a nucleus.  The finite size of the nucleus introduces a distance scale
$\Delta x$, which will be used with the uncertainty principle to bound the
energy loss.

\medskip
Components of momentum $q$ exchanged with the source can be related
to the source spatial size $\Delta x$ by  $\Delta x \Delta q\ge 1$.
 The ``$-$'' component of the exchanged momentum is easily probed.
Letting the four momentum squared of the final state be given by $M^2$,
then $M^2 = q^2 + 2p\cdot q$. We solve for $q^-\approx\Delta q^-\approx
M^2/(2p^+) - q^2/(2p^+)$, where $p^+= (E+p_z)/\sqrt{2}$

\medskip
Assume for definiteness now that we have two identical particles of mass $m$ in
the final state.  Their momenta in $(+, T, -
)$ notation can be listed as

\beqn
 p_{_1} &=& \left[ x_g E,\ \  \bar k_T,\ \  \bar k_T^2/ (2x_g E)
\right]\nonumber\\
 p_{_2} &=& \left[ (1-x_g) E,\ \  -\bar k_T,\ \  \bar k_T^2/ (2(1-x_g) E)
\right]
\eeqn
where $\bar k_T^2 = k_T^2 + m^2$ is the ``transverse mass".
The final state mass is
\beq
M^2 = \bar k_T^2\left({x_g\over 1-x_g} + {1-x_g\over x_g} +2\right)\ \ .
\eeq
 Assuming $\bar k_T^2<< M^2$, then the kinematics
forces either $x_g<<1$ or
$(1-x_g)<<1$. For gluons emitting gluons the two situations are
physically identical.  Thus it is a
good approximation to take

\beq
M^2 \approx {k_T^2+m^2\over x_g}
\eeq
Combining the above with $\Delta q^-=(M^2-q^2)/(2p^+)$, we have

\beq
1/ \Delta q^-\approx 2p^+ x_g/ (k_T^2+m^2-x_gq^2)\ \ ,
\eeq
This can be combined with the null plane uncertainty principle $\Delta
x^+ \Delta q^- \ge 1$ to give

\beq
2p^+ x_g/ (k_T^2+m^2-x_gq^2) < \Delta x^+ \ \ ,
\eeq

The meaning of $\Delta x^+$ is the light-cone ``time'' within which the
events occur.  This is clarified by a space-time cartoon (Fig 2).  From the
figure, due to the finite size of the nucleus, the $\Delta x^+ $ for events
causally propagating along the future light cone obeys  $\Delta x^+ < \sqrt{2}
L_A$, where $L_A$ is the rest frame length of the nucleus. Note that
non-relativistic propagation would allow $\Delta x^+ $ to become indefinitely
large if the parton stops inside the nucleus.  This possibility is irrelevant
to the discussion, although it would weaken the bound.

\medskip
BH did not include the
effects of $m^2$ and $q^2$. Ordinarily such terms are negligible in comparison
with large momentum transfers. However in this case we see that they are not
necessarily negligible in comparison with $k_T^2$. Setting $q^2$ and $m^2$
to zero in (8),
we obtain the result of BH
\beq
	x_g \le L_A k_T^2/(2E)
\eeq
using $p^+ = \sqrt{2}E$. The energy loss is then
given by $\Delta E_{\rm {parton}} \approx x_g E = x_g   x_1 E_p$, leading to
$\Delta x_1 = x_g x_1 \le L_Ak_T^2/2E_p$, as found by BH.

\medskip
Restoring the dependence on $m^2$ and $q^2$, we find:
\beq
x_g \le {(L_A/2E)(k_T^2 + m^2)\over 1-|q^2|L_A/(2E)}\ \ .
\eeq
At this point, if we know $|q^2|L_A/2E<<1$, then we have a BH bound with only
the modification of the ``transverse mass''. For our further discussion we will
assume this limit.

\bigskip
\noindent
{\bf 3.3 BACKUP}

\medskip
Are there any important loopholes in the bound?  We note the following:

\medskip
\noindent
i) The neglect of a final state mass for the parton is dangerous.  In the BH
calculation,  using $k_T $ values around the traditional values of 300 MeV,
then $k_T^2 = 0.1$ GeV$^2$, which is much smaller than any value one would use
for an effective $m^2$, even considering ``massless partons''.
The bound is quite
sensitive to this.  We will use $k_T^2$ values obtained from data.   This is
an important detail for application of the bound, but not for its concept.

\medskip
\noindent
ii) The bound, while invoking general kinematics and quantum principles,
nevertheless depends on the dynamics which assumed the production of
{\it two particles per vertex} in the final state.  It is similar in
spirit to electrodynamics, where the LPM
effect [8-10] provides the classic prototype for coherent
suppression of energy loss in a finite density target.

\medskip
To see this, we present a ``back of the envelop'' discussion of the physics of
LPM, abstracted from Feinberg and Pomeranchuk [10]. The kinematics assume
nearly
forward scattering of an electron (mass = $m_e$, energy =$E_e$) in a classical
medium, with emission of a bremmstrahlung photon carrying energy $\omega$. All
transverse momentum are assumed to be very small - this is an important point.
For forward scattering, one finds a spatial momentum transfer $q_{||}$ given by
\beq
q_{||} = \sqrt{E_e^2-m_e^2} - \sqrt{(E_e-\omega)^2-m_e^2}-\omega
\eeq
so that
\beq
\Delta q_{||} \approx m_e^2\omega/(2E_e^2)
\eeq
This momentum transfer determines a coordinate space region of longitudinal
length $r\sim 1/\Delta q_{||} \sim 2E_e^2/(m_e^2\omega)$. Suppose that the
length is so large that the electron undergoes several multiple scatterings.
{}From multiple scattering theory, the electron does a random walk with mean
square scattering angle
\beq
\theta^2_s\approx \left({E_s\over E}\right)^2{r\over L}
\eeq
where $E_s$ is an energy scale and $L$ a radiation length. Note that
$\theta_s^2$ is proportional to the distance travelled $r$. The LPM argument,
which is semiclassical, observes that coherence can be retained in
bremmstrahlung emission over a cone given by an angle set by the boost
 parameter
$\gamma=E/m$. Most photons are emitted with angle $\theta_\gamma\le 1/\gamma$
relative to the moving source. Coherence over the emission becomes crucial if
the multiple scattering angle $\theta_s$ is bigger than $\theta_\gamma$. This
condition is
\beq
\left({E_s\over E_e}\right)^2 {r\over L} \ge \left({m\over E_e}\right)^2\; ,
\eeq
which upon inserting $r\sim 2E_e^2/m_e^2\omega$ becomes
\beq
\omega\le {2E_e^2\over m_e^2}{E_s^2\over m_e^2 L}
\eeq
Emissions satisfying this bound are suppressed by destructive interference.

\medskip
The emphasis in the LPM analysis is on a region where the photon formation rate
is small compared to the collision rate. But then it follows that the process
is exquisitely sensitive to the scales setting the rate $\Gamma_{\rm form} =
k_T^2/2\omega$. By adjusting $k_T^2$ we can evidently scan across a range of
formation times, and turn the LPM suppression on or off. We will apply this
observation to the quarkonium suppression in
the next section.

\medskip
Continuing, the dynamics of QCD has features which are different from QED. In
QCD an incoming
gluon can split into three gluons at a single interaction due to the
perturbative 4-gluon vertex (Fig. (3)).  This upsets the QED--based argument.
We have  worked through the kinematics of three particles in the final state,
verifying that some regions reproduce the formation time argument while other
regions  exist which do not.

\medskip
For definiteness consider Fig. (3), in which an incoming gluon splits into
three gluons carrying momentum fractions $x_1$, $x_2$ and $x_3 = 1-x_1-x_2$.
The momenta are given by
$$P_1 = \left(x_1P,\ \ k_{T,1},\ \ {k_{T,1}^2\over 2x_1P}\right)$$
$$P_2 = \left(x_2P,\ \ k_{T,2},\ \ {k_{T,2}^2\over 2x_2P}\right)$$
$$P_3 = \left(x_3P,\ \ k_{T,3},\ \ {k_{T,3}^2\over 2x_3P}\right)$$
in $(+,T,-)$ notation, assuming $q^2$ and $q_T^2 \rightarrow 0$, and using
massless gluons. The $q^-$ momentum is simply given by the sum of the $p^-_i$:
$$
q^- = {k_{T,1}^2\over 2x_1P} +  {k_{T,2}^2\over 2x_2P} + {k_{T,3}^2\over 2x_3P}
$$
Applying $q^->1/L$, where $L$ is some interaction length, we can bound this
sum.

\medskip

The ``formation time'' is $\tau_{\rm form}=1/q^-$. We see that its inverse is
the
sum of three inverse formation times,
\beq
1/\tau_{\rm form} = 1/\tau_1 + 1/\tau_2 + 1/\tau_3
\eeq
where each formation time (up to trivial factors) equals the usual definition
{\it 1/}$\Gamma_i=\tau_i\sim 2x_iP/k_{Ti}^2$. The interpretation of (16) is of
course simpler in terms of rates: the total formation rate $\Gamma_{TOT}$
is the sum of the
three individual formation rates.

\medskip
 From the uncertainty principle the formation of the final state requires
$$\Gamma_{TOT}L\ge 1$$
or
\beq
L_A \ge {1\over \sum_i \Gamma_i}
\eeq
The finite value of $L_A$ means that not all the formation rates can be too
small or else they will destructively interfere.

\medskip
Unlike the case of 2-body formation - where the creation of one particle
implies the other - when we have a 4-point interaction, there is more than one
independent formation rate. In the relation (17) the biggest formation rate
wins.
That is, in the region
$$ {k_{T,1}^2\over 2x_1P}>> {k_{T,2}^2\over 2x_2P}+ {k_{T,3}^2\over 2x_3P}$$
then the creation of parton ``1'' dominates the issue of coherent
formation.
The
formation of partons ``2'' and ``3'', while occuring less rapidly, is triggered
by the formation of parton ``1'' at the 4-point vertex, {\it but they are not
separately resolved at this point}.

\medskip
Physically, here is what happens. An incoming gluon can be disrupted by the
source into a small $x$ gluon (say), while two larger-$x$ comoving gluons are
simultaneously created. The uncertainty principle applies to the smallest-$x$
gluon which cannot be
produced too slowly. But understanding the coherence and quantum mechanical
resolution of the smallest-$x$ gluon says {\em nothing about resolving two
other gluons into
separate components}. A subsequent hard collision (or similar independent time
scale) is needed to resolve them.

\medskip
Naturally we have a probe of a gluon's
momentum when a heavy quark is produced. Suppose, between
$x_2$ and $x_3$, we detect a quark carrying $x_2\approx x_F$; what fixes the
value of $x_3$? It
is fixed by detailed dynamics, not general principles.
This situation is unprecedented; it indicates the possibility of energy losses
not
following the rules of QED.
We conclude that the usual QED-LPM arguments are inadequate for a
quantitative analysis of energy loss in QCD.

\medskip
This is a real loophole; how big are its effects?   The four gluon
vertex is higher order in perturbation theory, being of order $g^2$, but
cannot be negligible because it is necessary for the gauge invariance of
the theory.\footnote{It has always been worrisome that the 4--gluon vertex
has produced few qualitative effects in high energy phenomenology.}  Moreover,
the problematic integration region is comparable or larger in
size to the region being discussed in the QED formation time arguments.  Yet
our
analysis of the dimensionless 4-gluon vertex emission does not introduce any
new ``large" scales beyond $L_A$, $k_T^2$, and $1/E$ already present.  By
dimensional analysis, then, something like the LPM analysis should survive
after
detailed integrations and combinatories over dimensionless quantities are
evaluated.
We will accept it for this study. We believe that further work could show that
the bound might be multiplied by a dimensionless factor we estimate to be a few
units.

\vglue 0.3in
\noindent
{\bf 4. APPLICATIONS}

\medskip
\noindent
We now turn to applying the energy loss bounds as practical tools.
The average value of $k_{_T}^2$ in our formula represents the transverse
momentum caused by scattering. We assume that this is the difference of
intrinsic transverse momentum $<k_T^2>_{\rm int}$, of the order of 0.5 - 1.0
GeV$^2$,
and the observed value. A more crisp definition can be given for ``leading
twist'' reactions but does not exist for power suppressed processes. To proceed
orderly we separate the initial state energy losses of gluons from final state
ones of heavy quarks. We present bounds based on initial state gluon energy
losses only. We will show that we produce a trend that is strikingly consistent
with the data. We discuss separating initial from final state effects in
Section 5.

\medskip
First we examine an analytic estimate of the energy loss. We assume gluons
are bremmed off with a distribution
$${dN\over dx_g} = f(x_g)\ ;\  f(x_g) = 0\ ,\ x_g>x_{\rm max} .$$
where $x_{\rm max}$ is given by the bound (9). Since we are in
quasi-nonperturbative region the details of $f_g$ are unknown. So long as
$f(x_g)$ is peaked at small $x_g$ but regular at $x_g = 0$ the details turn out
not to
matter much.

\medskip
The effective gluon distribution $\bar G(x)$ due to the shift from energy loss
is given by
$$\bar G(x) = \int_0^1 dx_gf(x_g) (1-x-x_g)^5$$
Using for example $f(x_g) = 1/x_g$, $x_{\rm min}<x_g<x_{\rm max}$, then we
are interested in the limit $x_{\rm max}<<1$. A series expansion can be
obtained
by integrating by parts
\begin{eqnarray*}
\int_{x_{\rm min}}^{x_{\rm max}}dx_g \left[{d\over dx_g} {\rm ln}x_g\right]
(1-x-x_g)^5 &=& {\rm ln}x_g (1-x-x_g)^5{\large|}_{x_{\rm min}}^{x_{\rm max}}\\
&+& 5\int_{x_{\rm min}}^{x_{\rm max}}dx_g{\rm ln}(x_g)(1-x-x_g)^4\ \ .
\end{eqnarray*}
The second term can be integrated by parts again leading to an asymptotic
series of any order desired. The first term is approximately
$$\bar G\sim {\rm ln}\left(x_{\rm max}/x_{\rm min}\right)(1-x)^5\left[1-
{5x_{\rm max}\over 1-x} + ...\right]$$
to first order in $x_{\rm max}$ and dropping terms proportional to $x_{\rm
min}$. The logarithm is
slowly varying in both $x_{\rm max}$ and the infrared cutoff $x_{\rm min}$
and will be ignored. One sees the GM mechanism emerging in the power series
expansion: the effects of a small $x_{max}$ get big as $x\rightarrow 1$.

\medskip
Now suppose that $\bar G$ is used in a calculation of cross section, namely
$${d\sigma\over dk_T^2dx_1dx_2} = x_1\bar G(x_1)\ x_2G(x_2){d\hat\sigma\over
dk_T^2dx_1dx_2}$$
where $d\hat\sigma$ is evaluated at the parton level. Setting up the $k_T$
integrals we have
 $${d\sigma\over dx_1dx_2} = \int d^2k_Tx_1 G(x_1)\ x_2G(x_2){d\hat\sigma\over
dk_T^2dx_1dx_2}\left(1-{5k_T^2L_A\over 2E(1-x_1)} + ...\right)$$
The second terms contains the nuclear effects; we have inserted the bound
$x_g=k_T^2L_A/2E$. Since the integrand is proportional
to $k_T^2$, we can do the integral to estimate the $A$-dependent correction as
$${1\over A}{d\sigma_A\over dx_1dx_2} \sim  {d\sigma_1\over dx_1dx_2}
\left(1-{5<k_T^2>_AL_A\over 2E(1-x_1)} + ...\right)\; .$$
The correction has a size set by $<k_T^2>_A$. For nuclei $L_A\sim 1.2$ Fm
$A^{1/3}$, and the data for $\Upsilon_{1s}$ production gives $<k_T^2>_A\sim
0.16
A^{1/3}$ GeV$^2 + <k_T^2>_D$. The kinematics of producing quarkonium with
invariant mass
$Q^2$ and momentum fraction $x_F$ requires $x_F = x_1 - x_2$, $Q^2/s = x_1x_2$,
which in the limit $Q^2/s<<1$ gives $x_F\approx x_1$. Then for the estimate for
$\Upsilon_{1s}$ production
$${1\over A}{d\sigma_A/dx_F\over d\sigma_1/dx_F} \approx  1-{5<k_T^2>_AL_A\over
2E(1-x_F)}\ .$$ This crude estimate does surprisingly well. Take for example
$A=184$ for tungsten,
and $x_F=0.5$. Then the effects of including energy loss lead to about 60\% of
the events compared to a
calculation neglecting energy loss.

To check the analytic estimate we did some
numerical calculations.
We wish to compare the experimental trends
in the data's x-dependence with its $k_T$ dependence. We will use the
experimental value of $<k_{_T}^2>_A - <k_{_T}^2>_{\rm int}$, where
$<k^2_T>_{\rm int}$ is the intrinsic value,
to calculate $\Delta x_1$. This can
then be used to calculate numerically the shift in the gluon distribution
functions due
to energy loss, thereby yielding the $x$ dependence in the
nuclear medium. Thus we take from the data
\beq
 \Delta x_1 \le L_A (<k_T^2>_A - <k_{_T}^2>_{\rm int})/(2E_p)\ \ .
\eeq
where we set $<k_{_T}^2>_{\rm int}$ to be equal to 0.91 GeV$^2$ in analogy to
Drell Yan [18].

	The x-integrated $k_T$ distribution of the quarkonia
has been parametrized experimentally [1]  by,
\beq
f_A(k_T^2) = \xi(A) \left[{1\over 1+(k_T/p_0)^2}\right]^6
\eeq
where $\xi(A)$ is an $A$ dependent normalization factor and $p_0$ is also
$A$ dependent.   The average value of $p_T^2$, defined by,
\beq
<k_T^2>_A = { \int_0^\infty dk_T^2 k_T^2 f_A(k_T^2)\over  \int_0^\infty
dk_T^2  f_A(k_T^2)}\ \ ,
\eeq
is equal to $p_0^2/4$. As given in Ref. [1], the values of $p_0$ for $^2$H
are

\centerline{$p_0$ = 2.78 for Drell Yan ,}

\centerline{$p_0$ = 3.22 for $\Upsilon$ .}

\noindent

\medskip
As discussed in the introduction, the values of $p_0$ are unfortunately not
available for
charmonium. The details of the $p_T^2$ dependence are not our object
here, and we would prefer to insert them from data.  To proceed with charmonium
we will
assume that it has the same transverse momentum dependence
as bottomonium-an assumption which can be relaxed when data is
obtained.  The value of $<k_T^2>_A$ for the case of bottomonium increases as
$0.16 A^{1/3}$ [14].  The experimental fit to this data for the case of
$\Upsilon$ is given in eq. (3).

We next need the production cross section in terms of the parton
distributions.  We assume that gluons give the most important contribution
to quarkonium production [19] for moderately large $x_f$ - this is discussed in
detail later. The cross section integrated over transverse momentum is then
given by

\beq
{d\sigma \over dQ^2dx_F} = {x_1x_2\over x_1+x_2} G(x_1+\Delta
x_1)G(x_2){\sigma^{gg\rightarrow c\bar c}(Q^2)\over Q^2}\ .
\eeq

\medskip
We apply the same procedure to the Drell Yan continuum dilepton data,
substituting quark and anti-quark distributions for the gluon
distributions.  The transverse momentum is quite different, but has been
re-fit to match the data.  The overall normalization is not relevant
because we report ratios of nuclear targets to the proton.  This procedure,
repeated for each nucleus,  gives us definite predictions for the x-
dependence of the experimental data, which is discussed in the next
section.

\medskip
\noindent
{\bf 5. RESULTS}

\medskip
In this section we discuss the results of our simple GM energy loss
calculation combined with the modified BH rule. The $x_F$ dependence of
the ratio $(d\sigma_A/dx_F)/d\sigma_D/dx_F)$ extracted by using the
experimentally measured $k_T$
dependence is shown in Figs. (4-7) for the cases of Drell Yan,
$\Upsilon_{1s}$ and charmonium respectively. For the bound we set $\Delta x_1$
(Eq. 18) to its maximum allowed value. We note that the
experimentally measured points are well within this theoretical limit; as
mentioned earlier, the trend in the data is to run parallel to the bound.
 For the case of charmonium we have taken the transverse momentum
distribution to be the same as for bottomonium. We see that this choice fits
the
charmonium $x_F$ dependence and therefore gives the minimum value of the
transverse momentum for the case of charmonium. The numerical results also
include shadowing besides energy loss. We assumed the following functional form
for the ratio $R_{\rm shadowing}$ of nuclear to deuteron quark distributions
due to
shadowing,
$$ R_{\rm shadowing} = 0.809 + 0.261\ {\rm exp}\left(-x_2-0.00526
A^{1/3}/x_2\right)$$
which fits the structure function data [20] with $\chi^2$/(degree of freedom) =
0.86.
The shadowing for the case of gluons is included by assuming that it is
the same as for quarks. The dashed lines in fig. (4), (5) and (6) represent the
results without including energy loss. We see that although the Drell-Yan
case is well reproduced by shadowing, charmonium and bottomonium data cannot be
explained purely be shadowing, but require the addition of gluon energy loss.

\medskip

Our results lead to several experimentally checkable predictions:

\vglue 0.3in
\noindent
{\it Small $k_t$}

\medskip

First, the conventional $A^\alpha$ analysis can be examined bin by bin in
$k_T$ and $x_F$.  Generally speaking, the energy loss picture is
distinguished by producing the largest suppression in the largest $k_T$
regions.  This does not mean that small $k_T$ is totally safe, because
rescattering can feed particles back i\rm \rm n to this region. However we
expect this
to be controllable and therefore predict
suppression increasing monotonically with $k_T$ at fixed $x_F$ or
integrated over  $x_F$.  This implies that the suppression should be reduced if
we consider the bin with $k_T$ less than about 1 GeV$^2$, where the dominant
contribution comes purely from the
primordial tranverse momentum. Our estimate of the $x_F$ distribution for
different $k_T$ cutoffs is given in Fig. (8). The perturbative part of
transverse momentum is again calculated by subtracting the intrinsic
contribution from the observed transverse momentum.
At low transverse momentum, $k_T^2<2<k_T^2>_{\rm int}$
we calculated the limiting value of $\Delta x_1$ by setting $k_T^2$ equal to
the intrinsic value of $0.91$ GeV$^2$.

The resulting curves, in Figure (8), show the ratios of the cross section as a
function of $x_F$ for various values of $k_T$. The curves are on a log plot,
because the overall normallization is not being predicted. A shift in
normalization $N$, translating the plots up or down, can be considered a free
parameter. Our object is the shape of the curves, which clearly evolves with
the
$k_T$ cut.
 For each region the experimental data should lie above the corresponding curve
and
follow the trend indicated in Fig. (8).

\vglue 0.3in
\noindent
{\it Large $x_F$}

\medskip

The idea of dominance of gluons in heavy quarkonia production is hardly new but
has been raised again recently by BQV [7].  They consider the lowest order
perturbative subprocess cross sections for producing a quark-antiquark
pair with invariant mass $Q^2$:

\noindent
{\it quark-antiquark channel:}
$$\hat\sigma^{q\bar q}(Q^2) = {2\over 9}\  {4\pi\alpha_s^2\over 3Q^2}(1+{1\over
2}\gamma)\sqrt{1-\gamma}\ ,$$

\noindent
{\it gluon channel:}
$$\hat\sigma^{gg}(Q^2) =
{\pi\alpha_s^2\over 3Q^2}\left[(1+\gamma+{1\over 16}\gamma^2){\rm
log}\left({1+\sqrt{1-\gamma}\over 1-\sqrt{1-\gamma}}\right) - \left({7\over 4}
+ {31\over 16}\gamma\right) \sqrt{1-\gamma}\right]\ ,$$

\noindent where $\gamma = 4 m_c^2/Q^2$.
Convoluting these cross sections with standard parton distributions, BQV
claim that the quark initiated process dominates over the glue-glue one
for $x_F \stackrel{>}{\sim} 0.5$.
 This conclusion is based on the perturbative
normalizations given above, and the fact that the quark distributions fall
less rapidly with $x$ than the gluons.
Quarks are forced kinematically to dominate as $x_F$ goes to 1.

	We agree with this in principle, but disagree that the crossover
point can be given by the Born term calculation.  Long experience
with detailed calculations of quarkonium production at high
energies favors gluons over quarks, indicating phenomenologically that the
perturbative normalizations are not to be trusted too literally. As already
noted, the data does
not allow the option of ascribing the quarkonium suppression to quark
channels while simultaneously accommodating the dilepton continuum.
Moreover, the Born term does not even give the Drell Yan cross section
correctly; for a long time it has been known that a ``K-factor" of
about 2 is needed to fit the data. Current understanding of K-factors is that
they summarize higher order corrections from initial and or final state
interactions.  One cannot assume the K-factors cancel out in nuclear
ratios: what is relevant is the relative amount of quark and gluon
contribution in each target.

	There are several ways to proceed. One can estimate the crossover
between quark versus gluon dominated quarkonium production with the K-
factor method. Using Ref. [21], in the initial state
interaction between two gluons we find a K-factor which is bigger than
the annihilation K-factor by $N_c/C_f$ = 9/4. With this (crude) estimate,
the crossover point for
quark annihilation channels over gluon channels at 800 GeV can be
estimated to be 0.65. One may also treat this as an adjustable parameter
to be determined experimentally once more data becomes available.

The value of large $x_F$ is that one can experimentally ``tune" the production
process to favor quark initiated reactions. We have already noted that the
Drell-Yan data indicates almost negligible energy loss, and smaller transverse
momentum, for light quarks compared to gluons.   If
this is correct, then as $x_F$ is increased above the crossover point the
suppression in the nuclear medium should diminish. In Fig (9) we present a
calculation illustrating the effect. This calculation was performed by using
the Eichten {\it et al.} [22] parametrization of the quark and gluon
distributions.
If the final state effects are negligible than the data should show a
sudden change in the current trend and an enhancement at large $x_F$.   This is
a dramatic signal meriting a careful search.  For this
calculation, we used the Born term cross sections modified by a relative
normalization of 9/4 for the gluons to the quarks.

Our approach has combined theory with empirical patterns taken from the data.
One could ask why light quarks do not come close to saturating the energy loss
bound while apparently gluons do. The answer is, we don't pretend to know. In
the same vein, one can ask whether final state heavy quark energy losses should
have been included. The answer is, the data does not indicate that significant
energy loss from the heavy quarks needs to be introduced.
 Nevertheless, toward
developing a truly model indpendent procedure, let us observe that the limit
$x_F\longrightarrow 1$ plays a key role. Suppose the up--turn as
$x_F\longrightarrow 1$ predicted above does not occur, even at such large
values of $x_F$ that we know the production is quark dominated. Then the
suppression of heavy quarkonium compared to Drell--Yan production {\it must} be
due to the heavy quark interactions above.  Similarly, comparing experiments
with different beams --
and especially pion beams known to be richer in quarks as $x\longrightarrow 1$
-- has the potential to help separate the final state from the initial state
effects.  The Badier {\it et al.} data on pion initiated
reactions [2] is consistent with this trend. We recommend using the data itself
to separate the issues in a systematic way.

\medskip
Certainly the effects discussed here have a direct bearing on
the use of charmonium or quarkonium suppression as a probe of quark-
gluon plasma formation at RHIC. Certainly a more thorough theoretical
understanding and further experimental investigation of the phenomena is
required before firm conclusions could be drawn from quarkonium production in
heavy
ion collisions.

\bigskip
\noindent
{\bf Note Added:}  After this work was completed we became aware
of a recent paper by M. S. Kowitt et al., Phys. Rev. Lett. {\bf 72}, 1318
(1994), which has extended the experimental
data on $x_F$ to larger values then was available previously. Their
results show that the ratio of nuclear to Deuterium production starts to rise
up considerably beyond $x_F=0.65$ but then falls again around $x_F=0.9$.
Except for the point at $x_F=0.9$, this data seems to
support our picture.
Comparing the data to our fig. (9), this suggests the idea that both light and
heavy
quarks lose negligible energy compared to gluons
may be correct.

\bigskip
\noindent
{\bf Acknowledgements:}  This work was supported in part under Department of
Energy Grant Nos. DE-FGO2-85ER-40214 and DE-FG05-91ER-40636  and the {\it
Kansas
Institute for
Theoretical and Computational Science}.

\newpage
\centerline{\bf Figure Captions}

\medskip
\begin{itemize}
\item[Fig. 1] Energy loss by emission of a parton in the presence of a source,
marked by a circled ``x".
\item[Fig. 2] Space--time picture of the null plane uncertainty principle
$\Delta x^+ \Delta q^-\geq 1$ The light--front time interval $\Delta x^+$ is
bounded to be less than about $\sqrt{2} L_A$ for ultra--relativistic processes.
\item[Fig. 3] Energy loss in a theory with a fundamental 4--point vertex; the
momentum fractions are indicated by the lengths of the lines. The uncertainty
principle bounds the formation rate of the fastest forming parton
(formation rate $k_T^2/2x_1P$) but says nothing further about the loss
occurring among the
other two partons, provided the sum of their  formation rates
$\Gamma_i=k_T^2/2x_iP$ is smaller than
the first.
\item[Fig. 4a-d] Ratio of the nuclear to Deuterium cross section for Drell-Yan
continuum dileptons (DY) as a function of $x_f$ calculated by including the
contribution only due to shadowing (dashed curve) and due to shadowing and the
maximum allowed value of energy loss (solid curve) for A=12, 40, 56 and 184,
respectively.
\item[Fig. 5] The parameter $\alpha$ for bottomonium as a function of $x_f$
calculated by including the
contribution only due to shadowing (dashed curve) and due to shadowing and the
maximum allowed value of energy loss (solid curve).
\item[Fig. 6a-d] Ratio of the nuclear to Deuterium cross section for $J/\psi$
production
as a function of $x_f$ calculated by including the
contribution only due to shadowing (dashed curve) and due to shadowing and the
maximum allowed value of energy loss (solid curve) for A=12, 40, 56 and 184,
respectively, using the transverse momentum distribution observed in the case
of bottomonium.
\item[Fig. 7] Predictions for the limiting values of ratios of nuclear to
Deuterium cross sections for $\Upsilon_{15}$ production as a function of $x_F$.
The data should lie above the
curves.
\item[Fig. 8] Limiting values of the ratio of Tungsten to Deuterium cross
section for $J/\psi$ production as a function of $x_f$ for different transverse
momentum
bins. The slopes of the curves at small transverse momentum are
much smaller than
the slopes at higher transverse momentum. Data in each
transverse momentum bin should have a slope less than or equal to that of the
plotted
curve. The overall $x_f$ independent normalization factor $N$ is  not
predicted.
\item[Fig. 9] Ratio of Tungsten to Deuterium cross section for $J/\psi$
production
as a
function of $x_f$ including contributions of gluon fusion and
quark-antiquark annihilation channels. The quark-antiquark annihilation
contribution overtakes charmonium production for $x_f>0.65$, producing the
upturn in the curves.
\end{itemize}

\newpage
\noindent
{\bf References}

\medskip
\begin{itemize}
\item[1.]  D. M. Alde et al., Phys. Rev. Lett. {\bf 66}, 133 (1991);  {\it
ibid} {\bf 66}, 2285 (1991); {\it ibid} {\bf 64}, 2479 (1990); M. J. Leitch et
al.,
Nucl. Phys. A{\bf 544}, 197c (1992).
\item[2.] J. Badier et al., Z. Phys. C {\bf 20}, 101 (1983).
\item[3.] S. Katsanevas, Phys. Rev. Lett. {\bf 60}, 2121 (1988);
\item[4.] S. J. Brodsky and A. H. Mueller, Phys. Lett. B {\bf 206}, 685 (1988);
A. Capella, J. A. Casado, C. Pajares, A. V. Ramallo and J. Tran Thanh Van,
Phys. Lett. B {\bf 206}, 354 (1988); J.-P. Blaizot and J.-Y. Ollitrault, Phys.
Lett. B {\bf 217}, 392 (1989);
S. J. Brodsky and P. Hoyer, Phys. Rev. Lett. {\bf 63}, 1566 (1989);
A. Capella, C. Merino, J. Tran Thanh Van, C. Pajares and A. Ramallo
Phys. Lett. B {\bf 243}, 144 (1990);
R. Vogt, S. J. Brodsky and P. Hoyer, Nucl. Phys. B {\bf 360}, 67 (1991)  ;{\it
ibid}  {\bf 383}, 643 (1992); R. C. Hwa and L. Lesniak, Phys. Lett. B {\bf
285}, 11 (1992); M. A. Doncheski, M. B. Gay Ducati and F. Halzen, Phys. Rev. D
{\bf 49}, 1231 (1994); B. Z. Kopeliovich and F. Niedermayer, Dubna preprint
JINR-E2-84-834, (1984);
R. V. Gavai and R. M. Godbole, preprint no. TIFR-TH-93-57; S. V. Akulinichev,
preprint no. MKPH-T-94-1.
\item[5.] S. Gavin and J. Milana, Phys. Rev. Lett. {\bf 68}, 1834 (1992).
\item[6.] S. J. Brodsky and P. Hoyer, Phys. Lett. B {\bf 298}, 165 (1993).
\item[7.] C. J. Benesh, J. Qiu and J. P. Vary, preprint No. ISU-NP-93-15.
\item[8.] L. D. Landau and I. J. Pomeranchuk, Dokl. Akad. Nauk. SSSR {\bf 92},
535 (1953); {\bf 92}, 735 (1953). English versions of these papers are
available
in L. Landau, {\it The collected Papers of L. D. Landau}, Pergamon Press, 1965.
\item[9.] A. B. Migdal, Phys. Rev. {\bf 103}, 1811 (1956).
\item[10.] E. L. Feinberg and I. J. Pomeranchuk, Nuovo Cimento, Supplement to
Vol. {\bf 3}, 652 (1956).
\item[11.] M. Gyulassy and X.-N. Wang, preprint No. CU-TP-598, LBL-32682.
\item[12.] J. Moss, Nucl. Phys. A {\bf 525}, 285c (1991) and D. M. Alde et al.,
Phys. Rev. Lett. {\bf 66}, 133 (1991).
\item[13.] J. C. Peng, private communication.
\item[14.] J. C. Peng, talk delivered at the workshop ``Perspectives in High
Energy Strong Interaction Physics at Hadron Facilities'',
Los Alamos, June 1-5, (1993).
\item[15.] J. W. Cronin et al., Phys. Rev. {\bf D11}, 3105 (1975); L. Kluberg
et al., Phys. Rev. Lett. {\bf 38}, 670 (1977); R. L. McCarthy et al., Phys.
Rev.
Lett. {\bf 40}, 213 (1978); D. Antreasyan et al., Phys. Rev. {\bf D 19}, 764
(1979); H. Jostlein et al., Phys. Rev. {\bf D20}, 53 (1979); Y. B. Hsiung et
al., Phys. Rev. Lett. {\bf 55}, 457 (1985); R. Gomez et al., Phys. Rev. {\bf D
35}, 2736 (1987); H. Miettinen et al., Phys. Lett. {\bf B207}, 222 (1988).
\item[16.] M. D. Corcoran et al., Phys. Lett. {\bf B259}, 209 (1991); C.
Stewart
et al., Phys. Rev. {\bf D42}, 1385 (1990); D. A. Finley et al., Phys. Rev.
Lett.
{\bf 42}, 1031 (1979); Y. B. Hsuing et al., Phys. Rev. Lett. {\bf 55}, 457
(1985); K. Streets et al., Phys. Rev. Lett. {\bf 66}, 864 (1991); P. B. Straub
et. al., Phys. Rev. Lett. {\bf 68}, 452 (1992).
\item[17.] M. Luo, J. Qiu and G. Sterman, Phys. Lett. B {\bf 279}, 377 (1992).
\item[18.] R. D. Field, {\it Applications of Perturbative QCD}, Addison-Wesley
Publishing Company (1989).
\item[19.] M. Gluck, J. F. Owen and E. Reya, Phys. Rev. D {\bf 17}, 2324
(1978).
\item[20.] J. Ashman et al., Phys. Lett. B {\bf 202}, 603 (1988); P. Amaudruz
et al., Z. Phys. C
{\bf 51}, 387 (1991); J. Ashman et al., Z. Phys. C {\bf 57}, 211 (1993).
\item[21.] A. P. Contogouris, S. Papadoupoulos and J. P. Ralston, Phys. Rev.
D. {\bf 25}, 1280 (1982).
\item[22.] E. Eichten, I. Hinchliffe, K. Lane and C. Quigg, Rev. Mod. Phys.
{\bf 56}, 579 (1984).

\end{itemize}
\end{document}